# Problems of Inheritance at Java Inner Class

Sim-Hui Tee
Faculty of Creative Multimedia
Multimedia University
Cyberjaya, Malaysia
shtee@mmu.edu.my

*Abstract*—Single inheritance has been widely accepted in the current programming practice to avoid the complication that incurred by multiple inheritance. Single inheritance enhances the reusability of codes and eliminates the confusion of identical methods that possibly defined in two superclasses. However, the mechanism of inner class in Java potentially reintroduces the problems encountered by multiple inheritance. When the depth of Java inner class is increased, the problem becomes severe. This paper aims at exposing the problems of inheritance at the Java inner class. In addition, a measure is proposed to evaluate the potential problem of inheritance for Java inner class

## I. INTRODUCTION

Single inheritance is a mechanism that allows a class to only inherit from a single superclass (or base class) [13][15]. It is adopted in most of the fourth generation programming languages such as Java, C#.NET, and PHP [8]. The primary advantage of single inheritance over multiple inheritance is that it discards the possibility of inheriting multiple implementations of the same methods [1]. In addition, single inheritance enhances greater reusability [2] and maintainability [3] as compared to multiple inheritance. Notwithstanding the advantages of single inheritance, it is as hard to use appropriately as multiple inheritance.

Java programming allows single inheritance to be used at both outer and inner class level. In Java, inner classes are classes that declared within a regular class, named outer class [16]. Inner classes are used as helper classes for their outer class [1]. They are defined to perform tasks which are related to the functionality of their outer class. They can be declared at any level of scope [4], as long as they are declared within outer class.

The use of inheritance in Java becomes complicated when one or more inner classes inherit from superclass. There is a potential reintroduction of the problem that encountered by multiple inheritance, such as the potential confusion of polymorphic implementation of methods. The use of inheritance in Java inner classes potentially leads to the difficulty of program comprehension and maintenance. The problem springs from the structures of inner class and natures of inheritance. This paper aims at exposing and discussing these problems. In addition, a measure is proposed to evaluate the potential problem of inheritance for Java inner class.

## II. RELATED WORKS

Recognizing the importance of inheritance in software design, reusability and maintenance [9][11], there are few inheritance metrics proposed by scholars. The inheritance metrics aims at analyzing the dependency of classes in a software application.

Chidamber and Kemerer defined two design decisions which relate to the inheritance hierarchy [5]. The first inheritance metric is Depth of Inheritance. It is a metric that measures the depth of the class in the inheritance tree [5]. This metric measures the extent to which the class is influenced by the properties of its superclasses [5]. Number of Children, which is the second inheritance metric proposed by Chidamber and Kemerer, measures the potential impact on the subclasses [5]. These two metrics are used to evaluate the genealogy of a class. They are separated from measuring the internal structures of class. Hence, they are not applicable to measuring inner class. In the work of Chidamber and Kemerer [5], inner class inheritance problem was not aware of.

Brito and Carapuca [6] proposed three types of inheritance metrics, which are Total Progeny Count (TPC), Total Parent Count (TPAC) and Total Ascendancy Count (TAC). TPC counts the number of classes that inherit, both directly and indirectly, from a superclass. TPAC counts the number of superclasses from which a class inherits directly. TAC counts the number of superclasses from which a class inherits directly or indirectly. Brito's and Carapuca's metrics are separated from measuring the internal structures of class as well. Their metrics only aims for the regular outer classes. Brito and Carapuca did not discuss inner class inheritance problems.

Mayrand et al [14] developed an assessment technique for the understandability of the inheritance graph of software. Their technique is based on metrics and graphical representation of the inheritance graph. They applied their technique to assess the topology of object-oriented software by extracting the main tree on the inheritance graph [14]. However, Mayrand et al did not extend their technique to assess inner class inheritance.

Beyer et al had investigated the impact of inheritance on metrics that measure a class [7]. They applied the flattening function to large C++ systems to examine the impact of inheritance on the size, coupling and cohesion metrics [7]. Beyer et al's work [7] was the first of its kind in studying the relation between inheritance and class component from the perspective of class metrics. However, Beyer et al did not investigate the relation between inheritance and inner class.

Previous works demonstrate that the investigations on inheritance are focused on two aspects: (1) genealogy of outer class and; (2) functional relationship between inheritance and outer class. Java inner class inheritance has not received attention from scholars. It is a missing part in the investigation of inheritance. This missing part is quite vital in understanding the relationship between Java inner class and genealogy of outer class. It is because Java inner class is not a mere component of its outer class. It is a class which can be instantiated. Thus, the functionality and structure of inner classes are more complicated than methods and attributes, although they belong to the same outer class. Furthermore, Java inner class presents a more complicated issue as it can contain attributes, methods, and other inner classes as well.

Notwithstanding the recognition of the importance of inheritance in object-oriented software paradigm, there are reported problems that caused by the use of inheritance. Most inheritance problems spring from the problem of inappropriate use. Wilde et al. hold that inappropriate use of inheritance presents difficulties in software maintenance [10]. Inappropriate inheritance can make the program dependencies hard to be analyzed during software maintenance [10].

The reusability of classes will be improved in the presence of deep inheritance hierarchy. However, the software is harder to maintain with deep inheritance hierarchy. It is a trend that inheritance hierarchies are kept flattened, sacrifice reusability for the sake of easy maintenance and comprehension [5]. Nasseri et al. have supported the tendency of flattened inheritance hierarchies with their empirical study on the evolution of seven open-source systems [12].

Previous works converge on an agreement that deep inheritance hierarchies are unfavorable in practice because of the introduction of complexity into the software. However, the problem of inheritance at the inner classes is not well aware of. Our research identifies these problems and their causes. In addition, a measure is proposed to aid the software developers in evaluating the potential threat of Java inner class inheritance problem.

III. PROBLEMS OF INHERITANCE AT JAVA INNER CLASS

Inheritance can be used at inner and outer class. When inheritance is found at inner class, there is a likelihood of cyclic inheritance which is undetected by Java compiler. In Java, cyclic inheritance is not allowed if a class, be it inner or outer, attempts to inherits from itself. Cyclic inheritance is absurd conceptually because it implies that a class is its superclass and subclass at the same time. Use of inheritance at inner class potentially results in detected or undetected cyclic inheritance where an inner class attempts to inherit from itself through its outer class or other external class. Figure 1 illustrates the problem of cyclic inheritance which can be detected by Java compiler.

```
class A extends A{
  class InnerA{ }
}
class B{
  class InnerB extends InnerB{ }
}
class C extends C.InnerC{
  class InnerC{ }
}
```

Figure 1. Problems of cyclic inheritance that are detected by Java compiler

The three classes in Figure 1 lead to compiler error. Cyclic inheritance occurs at outer class *A* and inner class *InnerB* because they attempt to inherit from themselves. As for class *C*, cyclic inheritance occurs at outer class because it inherits from its inner class. In Figure 1, cyclic inheritance leads to absurdness of class relationship as it implies that a class is its ancestor and descendant. Apparently, cyclic inheritance is a bad design in inheritance hierarchy.

However, Java is unable to detect potential cyclic inheritance when inheritance is used at inner class to inherit from its outer class. Figure 2 illustrates this situation.

```
class TestIn1{
  private void doPrinting( ){
    System.out.println("TestIn1");
  }

  class TestIn2 extends TestIn1{
    void call( ){
      doPrinting( );
    }
  }
}
```

Figure 2. Undetected indirect cyclic inheritance

In Figure 2, the inner class *TestIn2* inherits from its outer class *TestIn1*. Since the outer class *TestIn1* contains inner class *TestIn2*, *TestIn2* which inherits *TestIn1* also inherits itself. Java compiler fails to detect the conceptual error because the inner class inherits from itself indirectly, which implies an indirect cyclic inheritance. The method *doPrinting( )* is a private method of the outer class. In Java, private methods and attributes are non-inheritable. The method caller *doPrinting( )* in inner class does not use the private method callee *doPrinting( )* as an inherited method, but refers to it as a private property of its outer class. The cyclic inheritance relationship at inner class makes the code harder to comprehend.

Indirect cyclic inheritance may occur when the inner class inherits another class which inherits the outer class of that inner class, as shown in Figure 5. Java compiler is unable to detect this kind of cyclic inheritance.

```
class A{
  class B extends C{
  }
}
class C extends A{
}
```

Figure 3. Indirect cyclic inheritance between inner class and external class.

## IV. MEASURING POTENTIAL PROBLEM OF INHERITANCE FOR INNER CLASS

Causes of the problem of inheritance at inner class are taken into account in developing a preventative measure. Each Java class is assigned a measure value that indicates the potentiality of inheritance problem at inner class level. We call this measure as Potential Inner Class Inheritance Problem (PICIP). PICIP is a measure that carries integer value point. Each cause of the inheritance problem (as enumerated in Table 2) is conferred 1 value point of PICIP. All the value points are summed up to yield a total PICIP value. The greater the total PICIP value for a class, the greater potentiality of inheritance problem may occur at that class. High total PICIP value also implies harder program maintenance and greater program complexity. Hence, low total PICIP value is preferable for Java class. The maximum total PICIP value is 6, corresponding to each of six causes of inner class inheritance problem. The minimum total PICIP value is 0, implying no threat of inheritance problem at inner class. Minimum total PICIP value may also obtained when a Java class does not contain any inner class.

Figure 4 is a regular Java class without inner class defined in the outer class.

```
public class Account{
//...code implementation
}
```

Figure 4. Java class without inner class.

Based on the *Account* class in Figure 4, Table 1 is constructed to compute the corresponding PICIP value. The occurrence of each cause of inner class inheritance problem will be recorded as 1 at the right column of Table 1. However, the absence of the cause will be recorded as 0.

## V. EXAMPLES OF PICIP VALUE COMPUTATION

In this section, examples are given to illustrate the computation of PICIP value for Java class.

TABLE 1. PICIP VALUE FOR *Account* CLASS

| Cause of inner class inheritance problem | Value |
|---|---|
| The superclass of inner class is its outer class. | 0 |
| The superclass of inner class is inheriting from the outer class of that inner class. | 0 |
| The name of inner class is same with external class. | 0 |
| Overriding methods found in the inner class where cyclic inheritance takes place. | 0 |
| Deep level of inner class (more than one level) | 0 |
| Inheritance at outer class and inner class. | 0 |

When the values for each cause are summed up in Table 1, total PICIP value is 0. It implies that the *Account* class has no threat of potential inner class inheritance problem. It is because there is no inner class defined in the outer class.

In Figure 5, *House* class contains inner classes; whereas "Bedroom" is an identical name for an inner class and an external class.

```
public class House{
 public class Bedroom{
   public class Attachedwashroom extends
        Bedroom{

   }
  }
 }

 public class Bedroom{
 }
```

Figure 5. Java class with inner classes defined

Based on the classes in Figure 5, Table 2 is constructed to compute the corresponding PICIP value for *House* class. The second outer class *Bedroom* has a PICIP value 0 because it does not contain any inner class. *House* class contains an immediate inner class *Bedroom*. *Bedroom* inner class contains an immediate inner class *Attachedwashroom*. On the same vein, *Bedroom* class is the outer class for *Attachedwashroom* class. Hence, there are two levels of inner classes in *House* class, which are *Bedroom* and *Attachedwashroom*.

TABLE 2. PICIP VALUE FOR *House* CLASS

| Cause of inner class inheritance problem | Value |
|---|---|
| The superclass of inner class is its outer class. | 1 |
| The superclass of inner class is inheriting from the outer class of that inner class. | 0 |
| The name of inner class is same with external class. | 1 |
| Overriding methods found in the inner class where cyclic inheritance takes place. | 0 |
| Deep level of inner class (more than one level) | 1 |
| Inheritance at outer class and inner class. | 0 |

The total PICIP value for House class is 3, which is moderate on the scale from 0 to 6. The superclass of inner class *Attachedwashroom* is its outer class *Bedroom*. This will cause the cyclic inheritance problem. Besides, the inner class *Bedroom* has the same name with the external class *Bedroom*. This phenomenon adds the complexity of inheritance due to the identical class name. Lastly, deep level of inner class is found in *House* class, where it contains two levels of inner classes, which are *Bedroom* and *Attachedwashroom* inner class. Deep level of inner class induces low class cohesion because it tends to include irrelevant components to the functionality of the root outer class.

The proposed PICIP value enables the software developers to identify specific inner class inheritance problem. The PICIP value in Table 2 suggests that there are three problems need to be attended by software developers in order to avoid inner class inheritance problem. The rectification should be made by software developers based on the causes of the problem as shown in Table 2. The threat of inner class inheritance problem is eliminated if total PICIP value 0 is yielded after the program rectification.

## VI. CONCLUSION AND FUTURE WORKS

The use of inheritance at Java inner classes potentially leads to the difficulty of program comprehension and maintenance. In this paper, inner class inheritance problems have been identified and elaborated. A measure, namely PICIP, is developed to measure the potential problem of inheritance for Java class. Software developers are encouraged to reduce total PICIP value by rectifying the spotted problems. A minimum total PICIP value is desired because it implies the absence of inner class inheritance problem.

Future works should focus on studying the impact of deep level inner class on the inheritance hierarchy. Design of deep level inner class is not favorable as it adds complexity to the class and inheritance hierarchy.

In addition, PICIP should be validated in our future works. Validated PICIP should then be extended application-wise in order to understand the impact of inner class inheritance problem on the software application as a whole